\documentclass[10pt,a4paper,twocolumn,english,nobalancelastpage,prl]{revtex4-1}
\usepackage[utf8]{inputenc}
\usepackage{color}
\usepackage{amsmath}
\usepackage{amssymb}
\usepackage{graphicx}
\usepackage{esint}
\usepackage{caption}

\makeatletter

\usepackage{amsfonts}
\usepackage{eurosym}
\usepackage{amsthm}
\usepackage{braket}

\makeatother

\usepackage{babel}

\begin{document}
\global\long\def\abs#1{\left| #1 \right| }
 \global\long\def\ket#1{\left| #1 \right\rangle }
 \global\long\def\bra#1{\left\langle #1 \right| }
 \global\long\def\half{\frac{1}{2}}
 \global\long\def\partder#1#2{\frac{\partial#1}{\partial#2}}
 \global\long\def\comm#1#2{\left[ #1 ,#2 \right] }
 \global\long\def\vp{\vec{p}}
 \global\long\def\vpp{\vec{p}\, ^{\prime}}
 \global\long\def\dt#1{\delta^{(3)}(#1 )}
 \global\long\def\Tr#1{\textrm{Tr}\left\{  #1 \right\}  }
 \global\long\def\Real#1{\mathrm{Re}\left\{  #1 \right\}  }
 \global\long\def\braket#1{\langle#1\rangle}
 \global\long\def\escp#1#2{\left\langle #1|#2\right\rangle }
 \global\long\def\elmma#1#2#3{\langle#1\mid#2\mid#3\rangle}

\title{Quantum walk on a cylinder}

\author{Luis A. Bru$^{(1)}$, Germán J. de Valcárcel$^{(2)}$, Giuseppe Di
Molfetta$^{(3,4)}$, Armando Pérez$^{(3)}$, Eugenio Roldán$^{(2)}$,
and Fernando Silva$^{(2)}$}

\affiliation{$^{(1)}$Optical and Quantum Communications Group, ITEAM Research Institute,
Universitat Politècnica de València, Camino de Vera s/n, 46022--València,
Spain\\
$^{(2)}$Departament d'Òptica, Universitat de València, Dr. Moliner 50, 46100--Burjassot,
Spain\\
$^{(3)}$Departamento de F{\`i}sica Te{\`o}rica and IFIC, Universidad de Valencia-CSIC, Dr. Moliner 50, 46100-Burjassot, Spain\\
$^{(4)}$Aix-Marseille Universit{\'e}, CNRS, Laboratoire d'Informatique Fondamentale, Marseille, France
}

\begin{abstract}
We consider the 2D alternate quantum walk on a cylinder. We concentrate
on the study of the motion along the open dimension, in the spirit
of looking at the closed coordinate as a small or ``hidden'' extra
dimension. If one starts from localized initial conditions on the
lattice, the dynamics of the quantum walk that is obtained after tracing
out the small dimension shows the contribution of several components,
which can be understood from the study of the dispersion relations
for this problem. In fact, these components originate from the contribution
of the possible values of the quasi-momentum in the closed dimension.
In the continuous space-time limit, the different components manifest
as a set of Dirac equations, with each quasi-momentum providing the
value of the corresponding mass. We briefly discuss the possible link
of these ideas to the simulation of high energy physical theories
that include extra dimensions. Finally, entanglement between the coin and spatial degrees of freedom is studied, showing that the entanglement entropy clearly overcomes the value reached with only one spatial dimension.
\end{abstract}


\maketitle

\section{Introduction}

Quantum walks \cite{kempe2003quantum,kendon2006quantum,konno2008quantum,venegas2012quantum}
(QW) refer to a variety of dynamical processes that are quantum analogues
of classical random walks. As for their classical counterpart, there
is a basic distinction between continuous-time and discrete-time (or
coined) QWs, depending on whether time is a continuous parameter \cite{mulken2011continuous}
or a discrete one \cite{aharonov1993quantum}, in which case a ``quantum
coin'' is ``tossed'' at every step in order to decide the next
state of the system. Also, both classical and QWs can evolve on a
continuous space or on a lattice. An obvious (by definition) difference
between classical and QWs is that quantum superpositions and interferences
are inherent to the latter, which is at the root of their usefulness
in quantum algorithmic and quantum information in general \cite{ambainis2001one,childs2009universal,lovett2010universal}.
But the interest on QWs goes beyond this, as they can be understood
as simulators of the Schrödinger \cite{GdeValc_NJP2010,hinarejos2013understanding}
and Dirac equations \cite{strauch2006relativistic,di2012discrete,di2013quantum}.
Let us finally notice that some types of QWs can be implemented in relevant platforms
such as cold atoms and optical networks (see \cite{Manouchehri:2013:PIQ:2566741,preiss2015strongly}).

Here we study the discrete coined QW (DTQW) on a cylindric two-dimensional
lattice with rectangular geometry (call it cyl-QW), namely $\mathbb{Z}\times\mathbb{Z}/Q$,
where $\mathbb{Z}/Q$ denotes the cyclic group of the integers, modulo
$Q$. We note that DTQWs on simplicial complexes, including
cylinders, have been introduced recently \cite{matsue2016quantum}.
Our motivations for studying cyl-QWs are quite different. On the one
hand, existing materials such as carbon nanotubes already have a cylindric
geometry (with hexagonal cells in this case), so that DTQWs might
capture some of the elementary physics of transport in these systems
as they actually do, to some extent, with graphene; in fact, 2D-DTQWs
also exhibit an energy spectrum with conical intersections \cite{roldan2013n,hinarejos2013understanding};
on the other hand, the potential fragility of a 1D lattice, in which
broken links forbid the walk to progress, is obviously bypassed in
a cyl-QW. Moreover, the parallelism of transport on a cylinder could
be more resistant to dissipation and decoherence, especially when
originated from point defects. Finally, there is an especially appealing
motivation for us to study the cyl-QW, which lies in its continuous
limit, where space and time behave, effectively, as continuous variables.
Continuous limits of QWs have been studied quite many times \cite{strauch2006relativistic, knight2004propagating,GdeValc_NJP2010}, showing
that QWs recover the Dirac equation under proper assumptions \cite{GdeValc_NJP2010,hinarejos2013understanding,strauch2006relativistic,di2012discrete,di2013quantum}.
From this perspective, the cyl-QW could help in modeling the effect of
closed dimensions (maybe compact unobserved dimensions) on Dirac particles.
As we show below in detail, in this continuous limit the existence
of an unobservable closed dimension manifests as a mass term in the
Dirac equation, a mass that depends on the (pseudo-)momentum of the
initial state along the cyclic dimension, a situation that reminds
the tower spectrum in Kaluza-Klein theories (see for example \cite{Witten1981,RubakovPhys.Usp.44:871-8932001;Usp.Fiz.Nauk171:913-9382001}).
Here, Dirac particles with different masses are just selected by the
value of the momentum along the cyclic dimension (conserved by the
QW evolution). Clearly this attribution of the origin of mass to the
excitation of different modes in closed microscopic geometries is
in very much the same spirit as in string theory and other theories
that are based on the assumption of extra dimensions.

The rest of the article is organized as follows: In Section \ref{sub:II_A}
the cyl-QW is formulated. Analytical and numerical results of the
dispersion relations will be presented in \ref{sub:II_B}.
Then in section \ref{sec:III} we analyze the entanglement properties
of the model by deriving an analytical expression for the reduced
density matrix in the long term limit, starting from a localized initial
state. In section \ref{sec:IV} we compute the continuous limit of
the cyl-QW. Our main conclusions are summarized in Section V.

\section{Quantum walk on a cylinder\label{sec:II}}

In this section we first define and characterize the alternate quantum
walk (AQW) in 2D propagating on a cylinder. The corresponding dispersion
relations (DR) are derived in \ref{sub:II_B}. Some numerical simulations
will be presented to confirm the predictions made by the DR. 

\subsection{Formulation\label{sub:II_A}}

The AQW, first introduced by Ambainis et al. in \cite{Ambainis:2005}
for the 2D case, is the simplest way to build higher dimensional QWs,
as it makes use of a single qubit to alternate directions, instead
of a 4-level internal state. Interestingly, it was later shown to
be equivalent to the well-known Grover Walk in 2D \cite{CdiF_PRL2011,CdiF_PRA2011,roldan2013n}
and generalized to $N$ dimensions in \cite{roldan2013n}, where
its dispersion relations were analyzed in detail. 

We consider the quantum walker moving on a 2D discrete cylindrical
lattice oriented along the infinite $x$ direction, with $y$ indicating
the direction on the closed dimension. The total Hilbert space $\mathcal{H}$
corresponding to this system can be written as the tensor product
$\mathcal{H}=\mathcal{H}_{w}\otimes\mathcal{H}_{s}$, where $\mathcal{H}_{w}$
is the Hilbert space associated to the spatial degrees of freedom
with basis states $\left\vert m,l\right\rangle $, $m\in\mathbb{Z}$
and $l\in\left[0,Q-1\right]$, so that $Q$ is the number of nodes on
the closed dimension. The two-dimensional Hilbert space $\mathcal{H}_{s}$
corresponds to the internal (or spin) degrees of freedom of the walker,
and is spanned by the states $\{\ket 1,\ket{-1}\}$. Altogether, the
basis states of $\mathcal{H}$ can be written as $\left\vert m,l;s\right\rangle =\left\vert m,l\right\rangle \otimes\left\vert s\right\rangle $,
$s=\pm1$. The state evolution from time step $j$ to time step $j+1$
is dictated by an unitary evolution operator $\hat{U}$, so that$\left\vert \psi\left(j+1\right)\right\rangle =\hat{U}\left\vert \psi\left(j\right)\right\rangle $.
For the AQW, this operator is defined as 
\begin{equation}
\hat{U}=\hat{S}_{y}\hat{C}_{y}\hat{S}_{x}\hat{C}_{x},\label{eq:AEQWop}
\end{equation}
being $\hat{S}_{i}$ the conditional displacement along axes $i=x,y$,
\begin{eqnarray}
\hat{S}_{x} & = & \sum_{l=0}^{Q-1}\sum_{m=-\infty}^{+\infty}\sum_{s=\pm1}\left|m+s,l;s\right>\left<m,l;s\right|,\\
\hat{S}_{y} & = & \sum_{l=0}^{Q-1}\sum_{m=-\infty}^{+\infty}\sum_{s=\pm1}\left|m,l+s\,(\text{mod }Q);s\right>\left<m,l;s\right|.
\end{eqnarray}
and $\hat{C}_{i}$ the coin operator acting on the qubit, which can
be generally written as

\begin{equation}
\hat{C}_{i}=\begin{pmatrix}e^{i\left(\alpha_{i}+\beta_{i}\right)}\cos\theta_{i} & e^{i\left(\alpha_{i}-\beta_{i}\right)}\sin\theta_{i}\\
e^{-i\left(\alpha_{i}-\beta_{i}\right)}\sin\theta_{i} & -e^{-i\left(\alpha_{i}+\beta_{i}\right)}\cos\theta_{i}
\end{pmatrix},\label{C}
\end{equation}
with, in general, different angles for $i=x,y$.

The above definition for the displacement operators is equivalent
to imposing periodic conditions on the wavefunction at site $(m,l)$
with spin component $s$, defined as $\psi_{m,l;s}(j)=\left<m,l;s\mid\psi(j)\right>$.
One can therefore extend the support of this function to the set $\left(m,l\right)\in\mathbb{Z}^{2}$,
subject to the condition 
\begin{equation}
\psi_{m,l+Q;s}(j)=\psi_{m,l;s}(j)\ \forall l,j.\label{eq:BC}
\end{equation}
The probability of finding the walker at point $(m,l)$ at time step
$j$, regardless of the spin state, is given by
\begin{equation}
P(m,l,j)=\sum_{s=\pm1}|\psi_{m,l;s}(j)|^{2}.
\end{equation}
We will be mostly concerned about the propagation of the walker along
the open dimension $x$, thus implicitly assuming that the closed
$y$ dimension is ``small'' as compared with the spread along the
tube. In other words, we treat the propagation along $y$ as unobservable,
and consider only the marginal probability 
\begin{equation}
P(m,j)=\sum_{l=0}^{Q-1}\sum_{s=\pm1}|\psi_{m,l;s}(j)|^{2}.\label{eq:marginalprob}
\end{equation}

\subsection{Dispersion relations\label{sub:II_B}}

\begin{figure}[h]
\includegraphics[width=0.80\columnwidth]{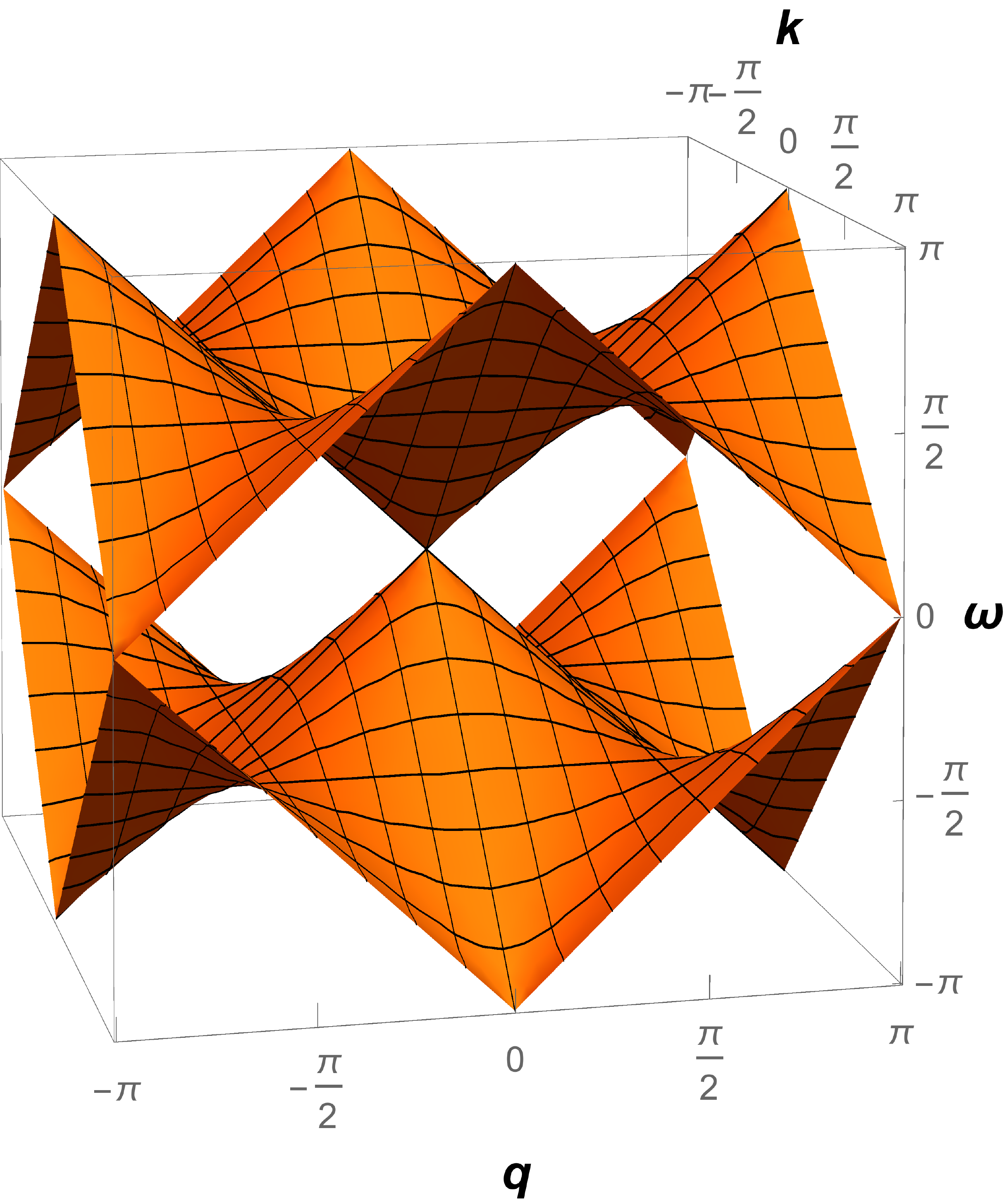}
\includegraphics[width=0.80\columnwidth]{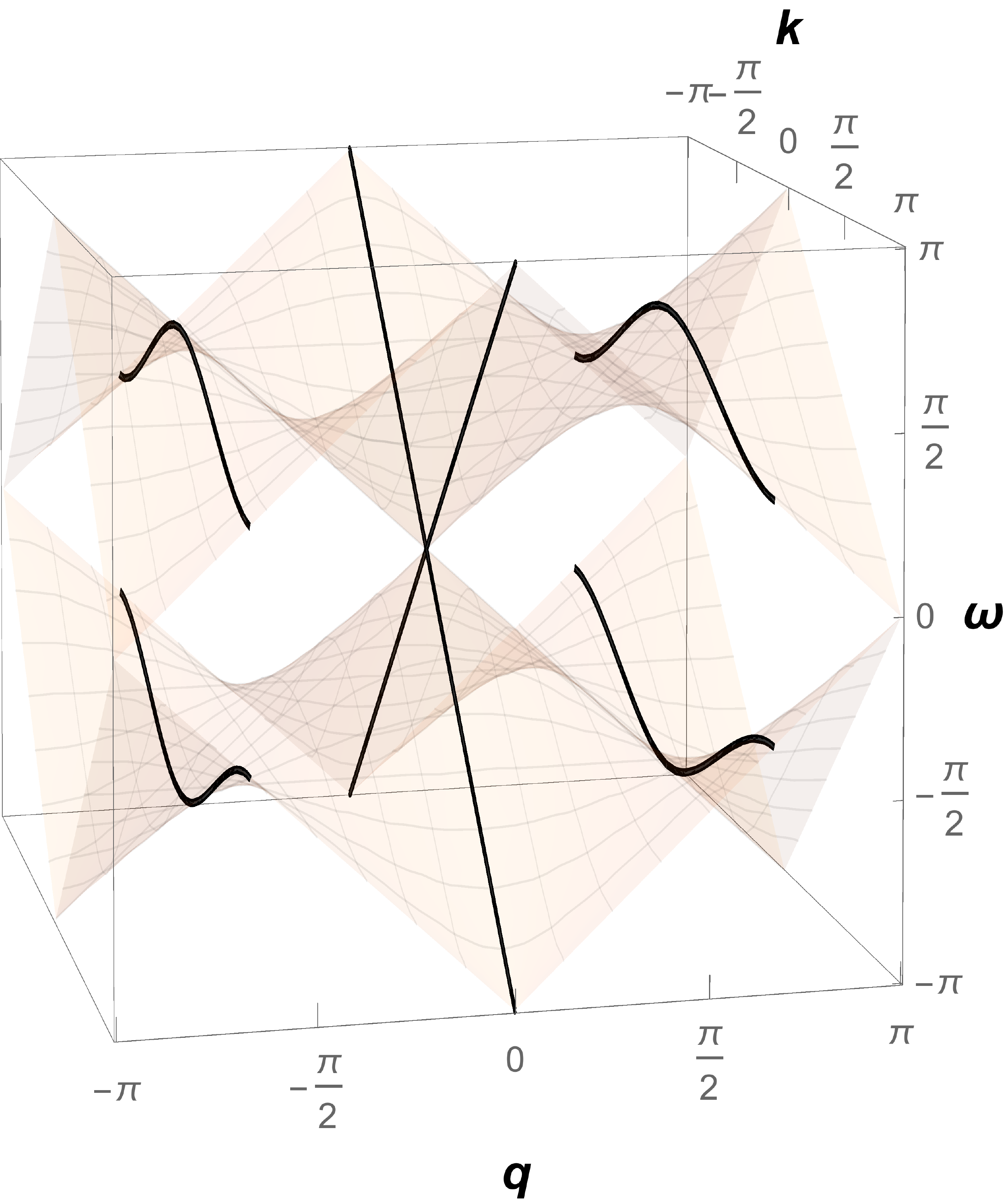}
\captionsetup{singlelinecheck=off,font=small, justification= centerlast}
\caption{(Color online)Dispersion relation of the AQW in 2D for the conventional case (a)
and for the cylinder with $Q=3$ nodes around cyclic dimension (b).
Both of them correspond to the Hadamard case $\theta_{x}=\theta_{y}=\pi/4$. }
\label{fig:DR_AQW}
\end{figure}

The spectrum of any QW is an essential tool to understand its behavior
\cite{Nayak2007}. Provided that the unitary operator is translationally
invariant, the system can be described in terms of quasi-momentum states
$\ket{k,q}$, where $k$ corresponds to the $x$ direction, and $q$
to the $y$ direction, respectively. Using this basis, the unitary
operator (\ref{eq:AEQWop}) adopts the expression
\begin{equation}
\hat{U}_{q}(k)=\begin{pmatrix}A_{11}&A_{12}\\A_{21}&A_{22},
\end{pmatrix}
\end{equation}
where 
\begin{eqnarray}
A_{11}=e^{-i(k+q)}\left(c_{x}c_{y}+e^{2ik}s_{x}s_{y}\right) \nonumber \\
A_{12}=e^{-i(k+q)}\left(c_{y}s_{x}-e^{2ik}c_{x}s_{y}\right)\nonumber\\
A_{21}=e^{-i(k-q)}\left(c_{x}s_{y}-e^{2ik}c_{y}s_{x}\right) \nonumber \\
A_{22}=e^{-i(k-q)}\left(e^{2ik}c_{x}c_{y}+s_{x}s_{y}\right),\nonumber
\end{eqnarray} 
with the notation $c_{i}=\cos\theta_{i}$, $s_{i}=\sin\theta_{i}$.
In the latter equation, we have set all phases to zero $\alpha_{i}=\beta_{i}=0$
for $i=x,y$, since the coin angles $\theta_{x}$ and $\theta_{y}$
are the only dynamically relevant parameters. One can readily obtain
the eigenvalues of $\hat{U}$, which can be written as $e^{i\omega_{\pm}}$,
where 
\begin{equation}
\cos\omega_{\pm}=c_{x}c_{y}\cos\left(k+q\right)+s_{x}s_{y}\cos\left(k-q\right)\label{eq:DR}
\end{equation}
defines the dispersion relations. The function $\omega_{\pm}\left(k,q\right)$
is $2\pi$-periodic along every component, due to the discreteness
of the lattice. This allows us to restrict ourselves to the first
Brillouin zone $k,q\in\left]-\pi,\pi\right]$. Fig. \ref{fig:DR_AQW}
shows the DR when the coin operators $\hat{C}_{i}$ are both chosen
to be the Hadamard coin, i.e. $\theta_{x}=\theta_{y}=\frac{\pi}{4}$.
In this twofold band structure, the most remarkable feature is the
presence of conical intersections, where the two bands meet. This
feature is strongly related to the propagation properties of the AQW:
conical intersections are present whenever $\theta_{x}=\theta_{y}$
and they establish a mean to swap population between bands. Pure linear
spreading takes place close to those points, When angles are different
$\theta_{x}\neq\theta_{y}$ the contact points disappear, thus avoiding
linear spreading and allowing zero group velocity eigenstates, thus
producing a dramatic localization of the wavefunction at the origin.
This key feature was recently used to build an electric QW in 2D with
almost perfect localization in \cite{LBru_PRA2016}. For further details
on the derivation of the DR and the influence of conical intersections
we refer the reader to \cite{roldan2013n,LBru_PRA2016}.

%\begin{figure}
%\begin{centering}
%\subfloat[\label{fig:DR_AQW}]{\centering{}\includegraphics[scale=0.25]{FIG/DR_1}
%}\par\end{centering}
%\begin{centering}
%\subfloat[\label{fig:DR_tube_3}]{\centering{}\includegraphics[scale=0.25]{FIG/DR_2}
%}
%\par\end{centering}
%\caption{Dispersion relation of the AQW in 2D for the conventional case (a)
%and for the cylinder with $Q=3$ nodes around cyclic dimension (b).
%Both of them correspond to the Hadamard case $\theta_{x}=\theta_{y}=\pi/4$.}
%\end{figure}

So far, we have considered the DR (\ref{eq:DR}) for arbitrary values
of $k$ and $q$. However, it is easy to show that cyclic conditions along
the closed direction (\ref{eq:BC}) restricts the possible values
of $q$ to the set 
\begin{equation}
\{q_{i}=\frac{2\pi i}{Q},\text{ where }i\in\mathbb{Z}\}.
\end{equation}

This set contains $Q$ different values in the first Brillouin
zone, with a distribution that depends on whether $Q$ is an odd or
an even number. This is due to the restrictions introduced by the DR symmetry properties. When $Q$ is even some degeneracies appear in the spectrum because of the symmetry of the DR; moreover, the number of degeneracies is different when $Q$ is multiple of 4 so that the number of different discrete states appearing in the spectrum is (Q+4)/4 when $Q$ is multiple of 4 and $(Q+2)/4$ when it is not. Importantly, when $Q$ is multiple of 4 some of the states in the spectrum become flat (see Fig. 2).

\begin{figure}[h!]
\includegraphics[width=0.90\columnwidth]{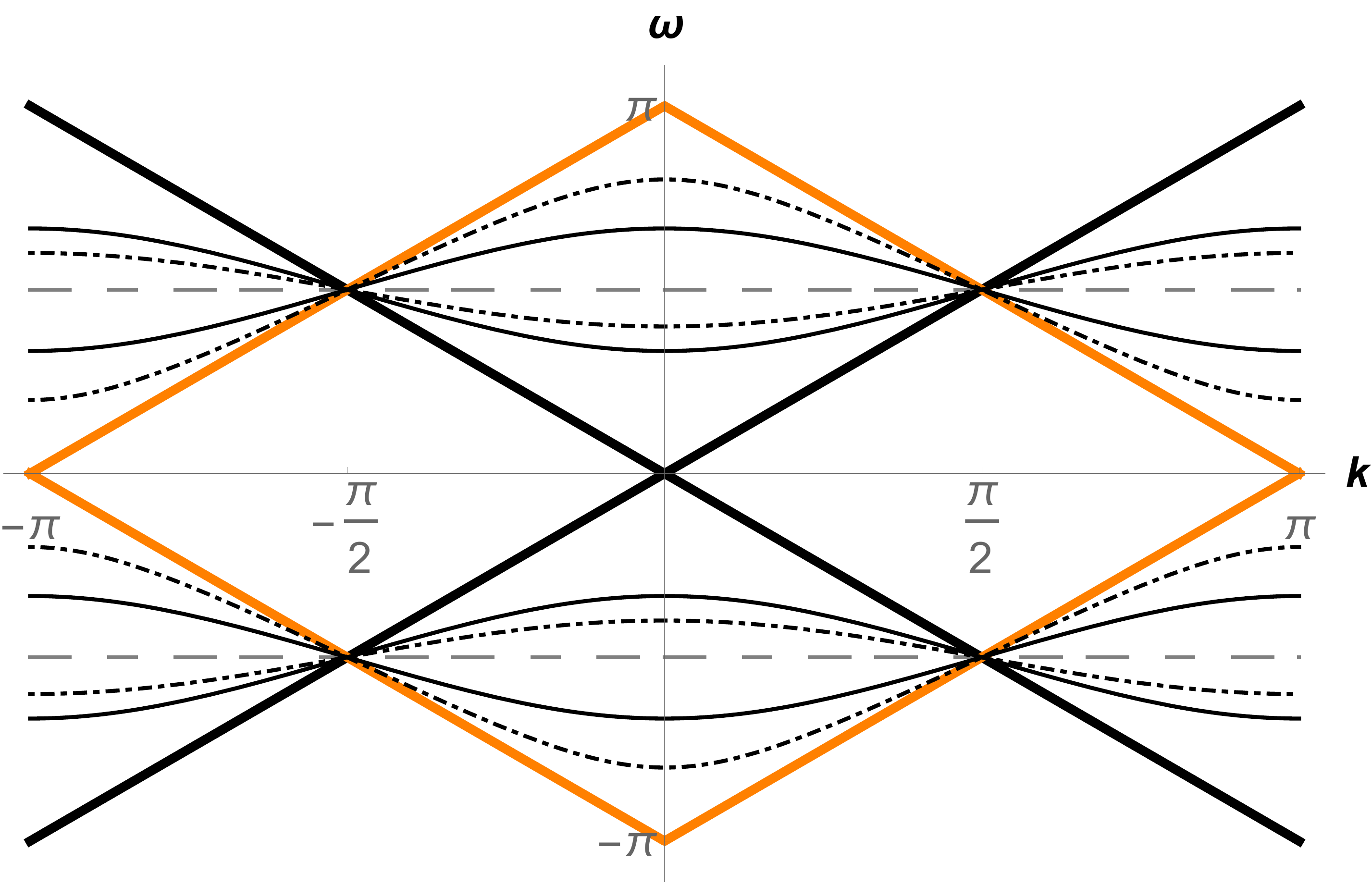}
\captionsetup{singlelinecheck=off,font=small, justification= centerlast}
\caption{(Color online) Dispersion relations of the AQW on the cylinder for the case $Q=4,5$ and 6, plotted along the quasi-momentum of the open dimension $x$. The thick line (orange and black) contributions are always present, but the orange one is only present for even cases $Q=4$ and 6. Thin solid lines correspond to $Q=6$, and the dot-dashed ones to $Q=5$. The dashed horizontal line only appears for $Q=4$.} \label{fig:DRs_Q456}
\end{figure}

\begin{figure}[h]
\includegraphics[width=1.\columnwidth]{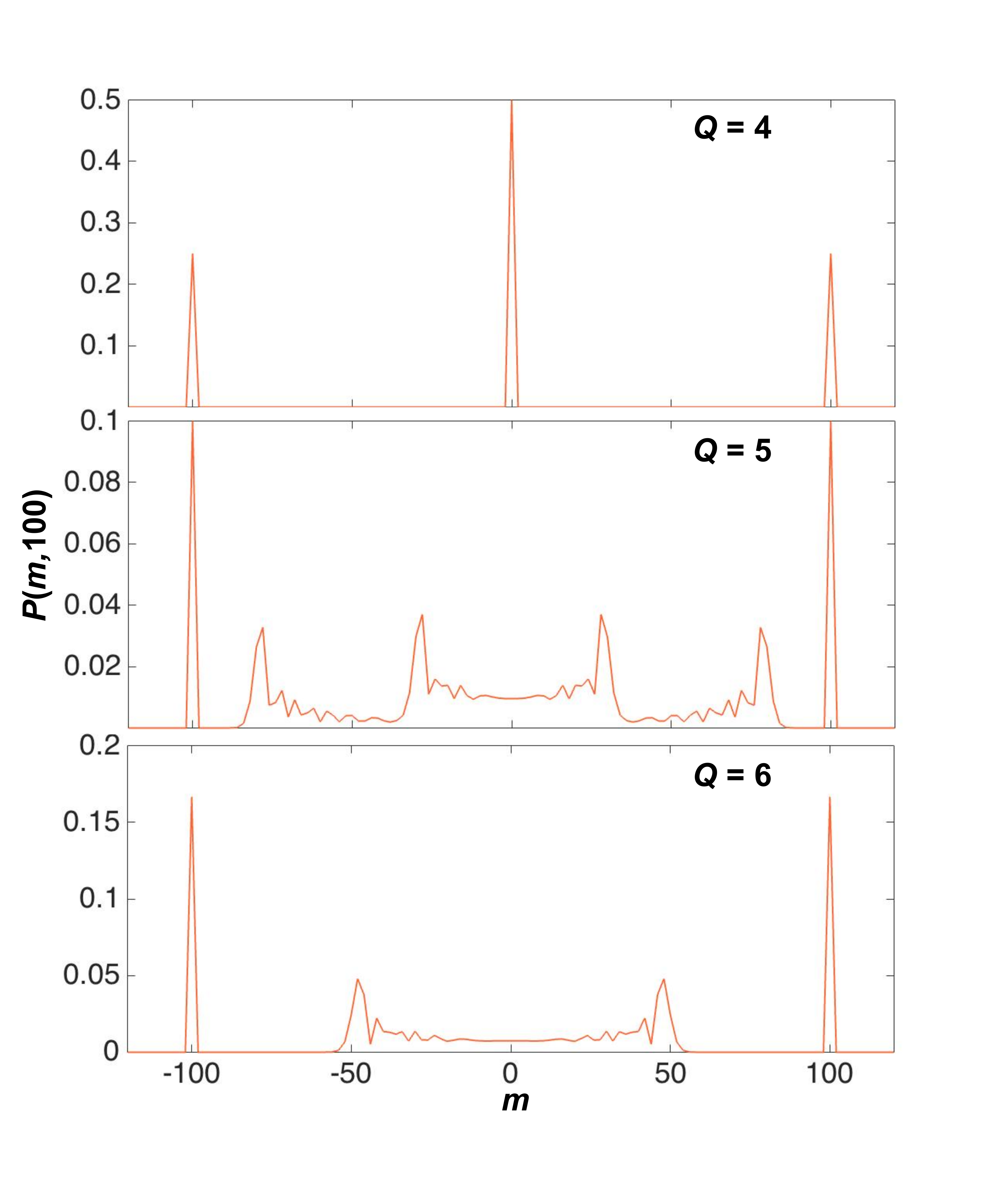}
\captionsetup{singlelinecheck=off,font=small, justification= centerlast}
\caption{(Color online) Marginal probability distribution,
as defined by Eq. (\ref{eq:marginalprob}) for an AQW on the cylinder
in the Hadamard case after $j$=100 steps, starting from a localized
initial condition $\left\vert \psi(0)\right\rangle =\frac{1}{\sqrt{2}}\left(1,i\right)$
at the lattice origin $(0,0)$. Different values of $Q=4,5$ and 6
have been represented, from top to bottom.}
\label{fig:Prob_dist}
\end{figure}

In order to interpret the derived DR, let us take a closer look to
the Hadamard case (we will restrict ourselves to this case in what
follows). Since we are mostly interested in the propagation along
the open dimension $x$, the dynamics is governed by $Q$ contributions
of the type: 
\begin{equation}
\cos\omega_{\pm}=\cos k\cos q_{i}.
\label{eq:DR}
\end{equation}
We notice that this formula represents a set of DRs of several 1D
quantum walks \cite{GdeValc_NJP2010}, with each $q_{i}$ playing
the role of the different $\theta$ angles of the coin operator. Therefore,
we expect the \textit{AQW on the cylinder to be described by a set
of several 1D QWs propagating along $x$}, with different propagation
velocities given by the corresponding maximum group velocities obtained
from $q_{i}$. This result is one of the major results of this work, and is confirmed by our numerical simulations in the following subsection. Moreover, we can select one or several of 1D QWs just preparing accurately the initial state in the momentum space, as in Fig. \ref{fig:Prob_dist}. \\ 

%In order to make an exemple consider the AQW on the cylinder shown in figure \ref{fig:DRs_Q456}.
%The localized initial condition is defined by $\left\vert \psi(0)\right\rangle =\frac{1}{\sqrt{2}}\left(1,i\right)$ at point $(0,0)$. Since the initial condition is completely localized in space, this implies the corresponding quasi-momentum wave function
%is uniformly distributed over all $\left(k,q\right)$ possible values, being also equally distributed between the two bands. 

%In Fig. \ref{fig:Prob_dist} we have plotted the marginal probability
%distribution, as defined by Eq. (\ref{eq:marginalprob}) after $j=100$
%time steps, for a cylinder with $Q=4,5$ and 6. 

Fig. \ref{fig:Prob_dist} confirms the agreement between our numerical simulations
and the predictions bases on the DR in each case. For
example, for $Q=6$ we have two different non-degenerated contributions,
one of them with maximum group velocity propagation. This type of
`massless' component, which is present in all cases, is stronger for
even than for odd values of $Q$ due to the lower number of total contributions.
The presence of this component avoids any possibility of localization
of the wave-function. As discussed above, this feature is ultimately
due to the presence of conical intersections in the present AQW system.

On the other hand, the existence of a strictly localized component
at the origin for $Q=4$ is due to the presence of a zero group velocity
1D contribution in the set. These $\omega$ flat contributions will
always be present whenever $Q$ is a multiple of 4. In all these cases,
localization at the origin will show up.

\section{Entanglement\label{sec:III}}

Entanglement between the coin and spatial degrees of freedom is generated
as a consequence of the evolution of the QW \cite{carneiro2005entanglement,venegas2005quantum,endrejat2005entanglement,abal2006quantum,omar2006quantum,maloyer2007decoherence,pathak2007quantum,liu2009one,annabestani2010asymptotic}.
The amount of entanglement can be quantified using the von Neumann
entropy of the reduced density matrix of the coin degrees of freedom,
after tracing out the spatial ones. More precisely, we define this
quantity, as a function of the time step $j$, by 
\begin{equation}
S(j)=-Tr\left\{ \rho_{c}(j)\log_{2}\rho_{c}(j)\right\} ,
\end{equation}
where $\rho_{c}(j)\equiv\sum\escp{m,l}{\psi(j)}\escp{\psi(j)}{m,l}$
is the reduced density matrix for the coin space. 
A measure of the entanglement entropy was first numerically obtained in \cite{carneiro2005entanglement}, and
proven later in \cite{abal2006quantum} that, for a Hadamard walk
with localized initial conditions the asymptotic entanglement is $S_{lim}\simeq0.8720$
for all initial coin states, although higher values can be reached
by starting from non-localized conditions (see also \cite{de2013multidimensional}).
The question that arises is whether the quantum walk on a cylinder
is also limited to this amount of entanglement, when the evolution
starts from a localized state. Our goal is to obtain an analytical
expression for the reduced density matrix $\rho_{c}(j)$ in the long
term limit. This calculation is more conveniently done in the quasi-momentum
space and detailed in the appendix \ref{sec:appendixA}. We consider an initial state localized at $m=0,l=0$ and
arbitrary coin components, i.e. 
\begin{equation}
\ket{\psi(0)}=\cos\frac{\theta}{2}\ket{0;0;1}+e^{i\phi}\sin\frac{\theta}{2}\ket{0;0;-1},
\end{equation}
where $\theta$ and $\phi$ represent the angles of the initial state
on the Bloch sphere. 

\begin{figure}[h]
\includegraphics[width=0.48\columnwidth]{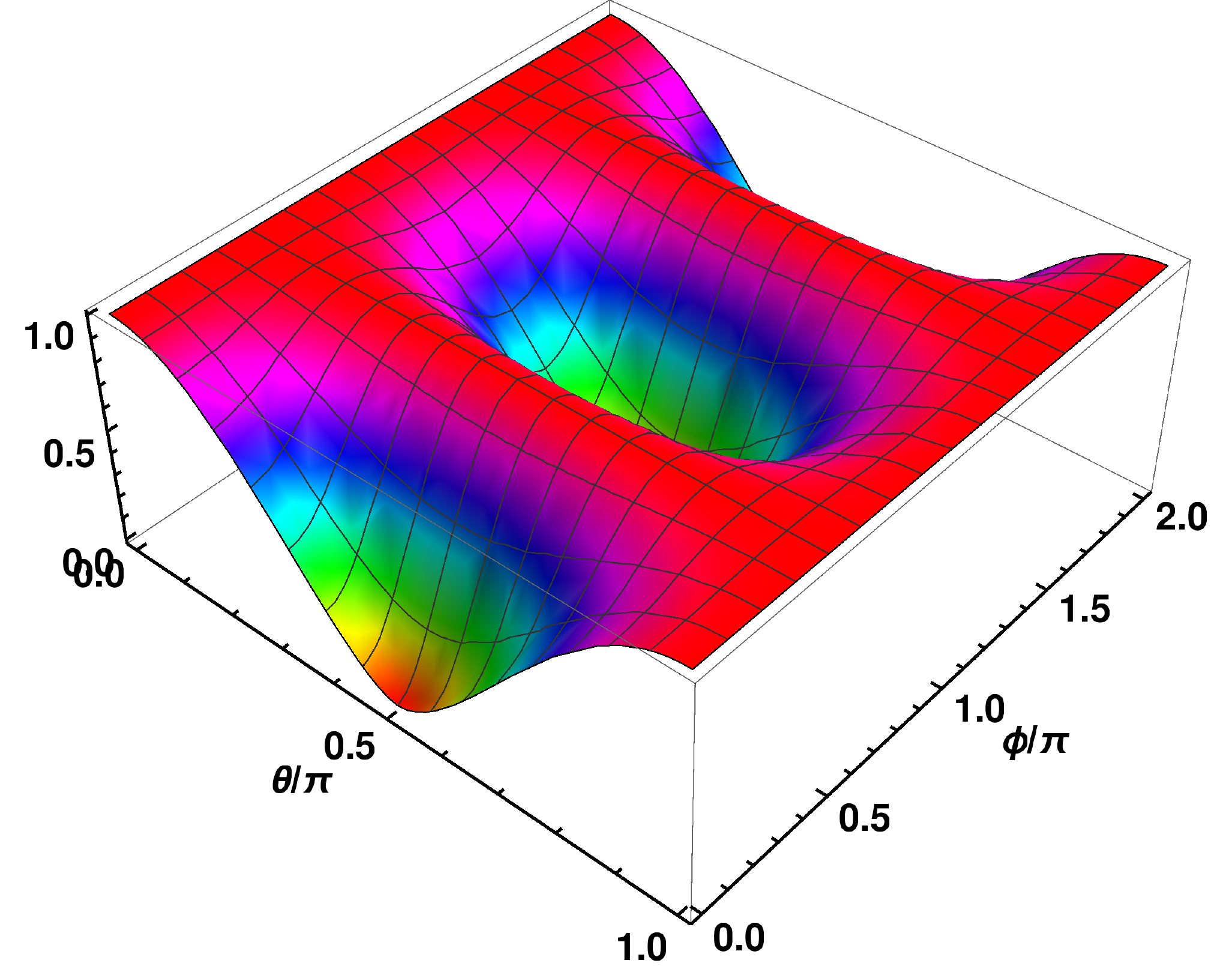}
\includegraphics[width=0.48\columnwidth]{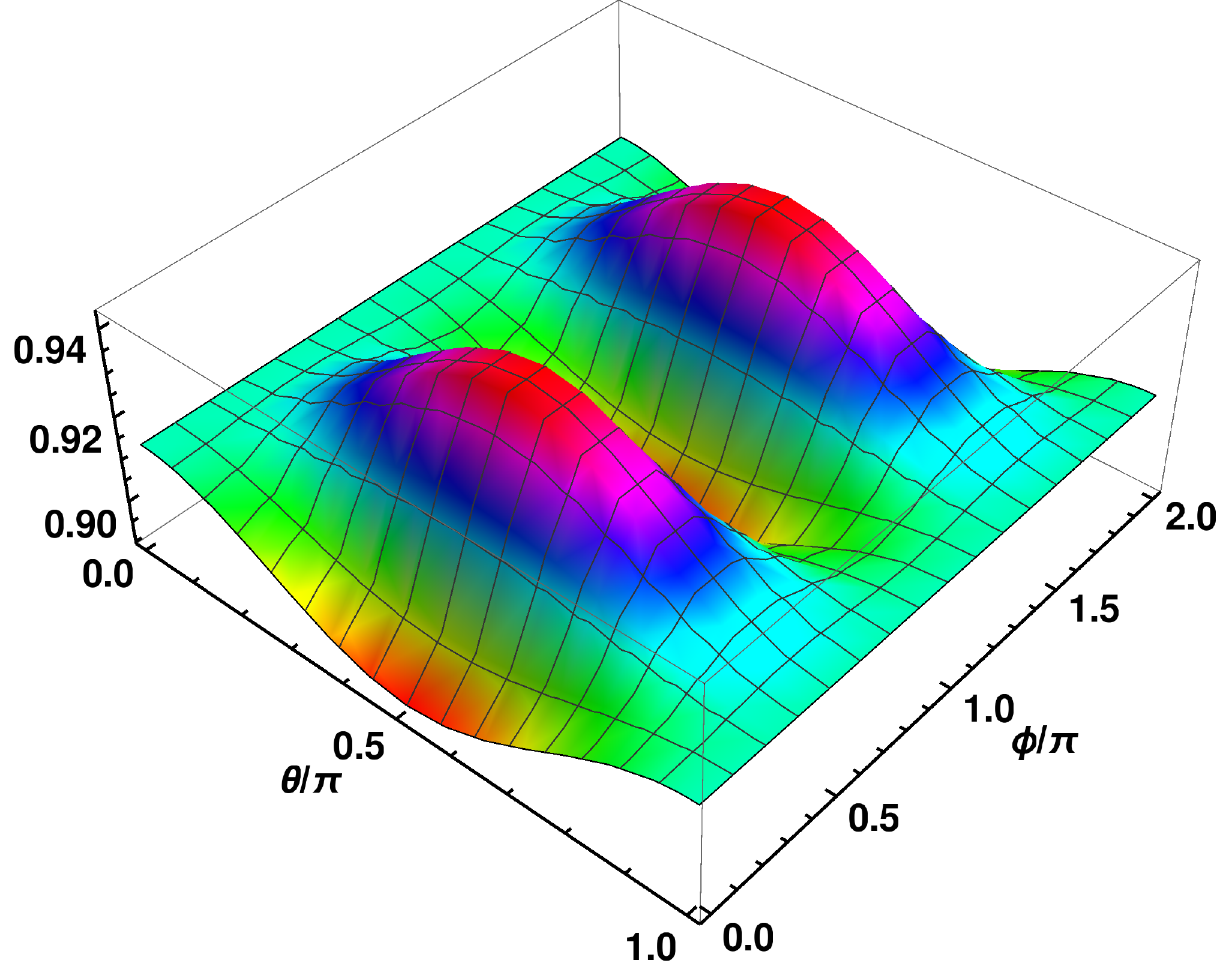}
\includegraphics[width=0.48\columnwidth]{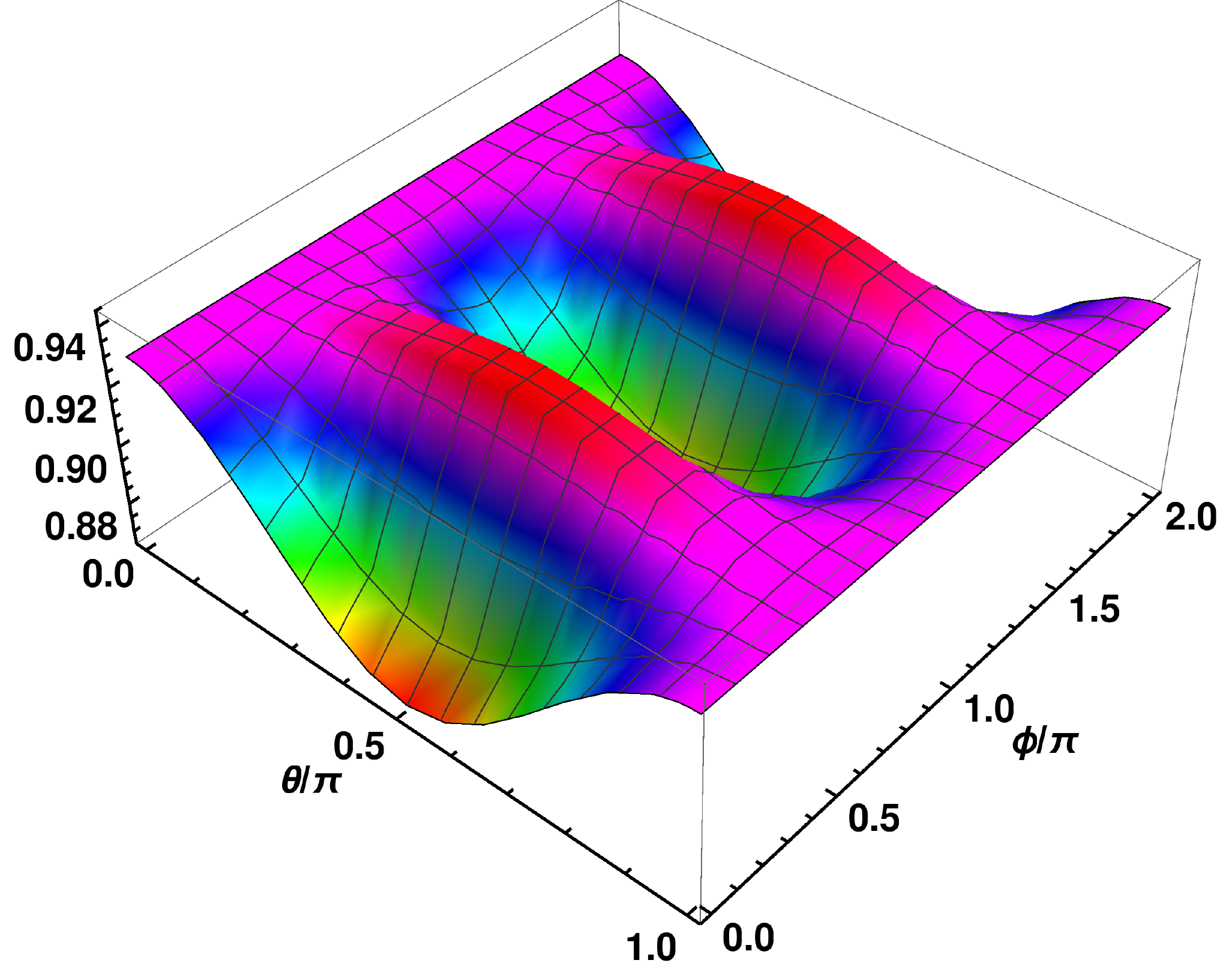}
\includegraphics[width=0.48\columnwidth]{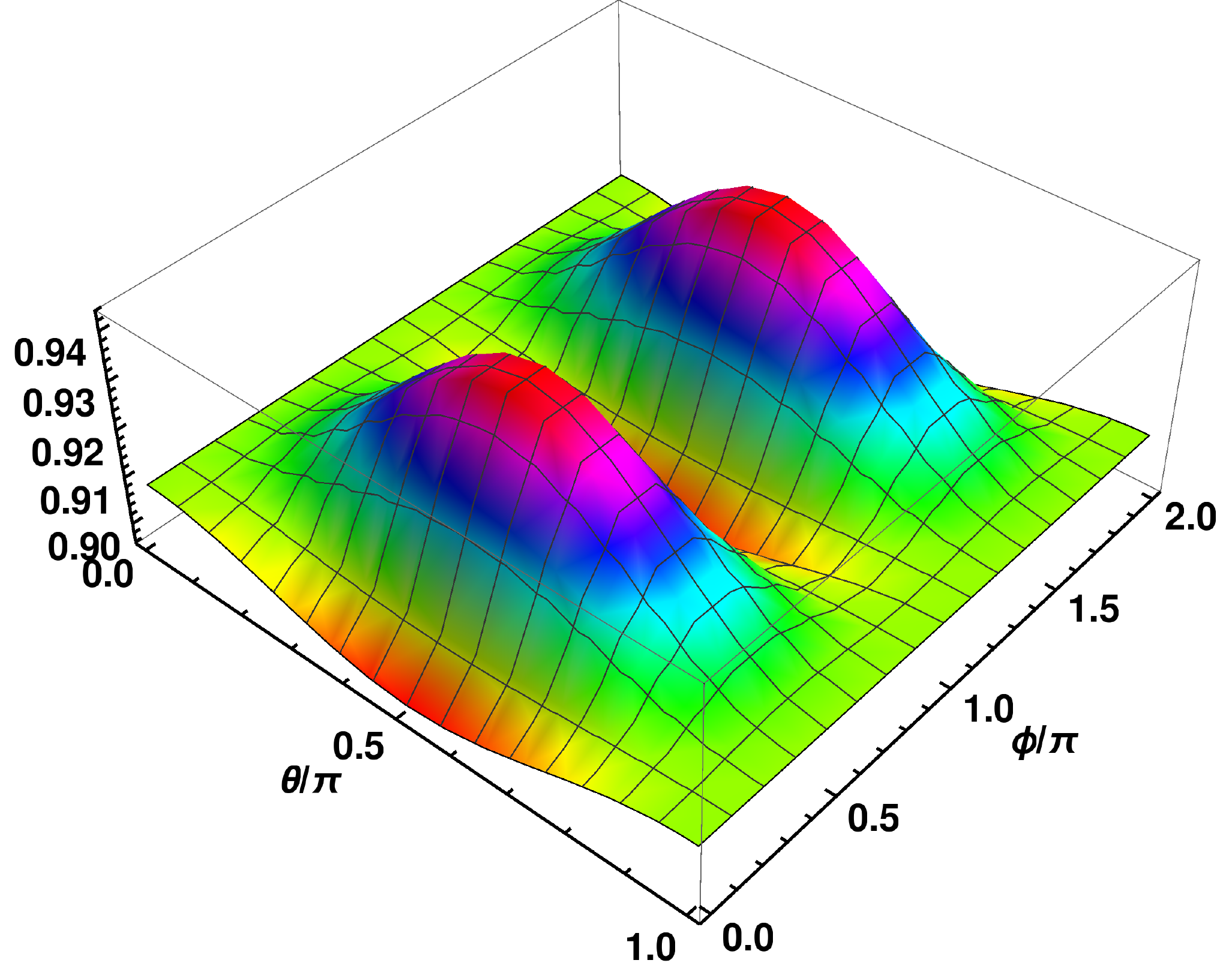}
\captionsetup{singlelinecheck=off,font=small, justification= centerlast}
\caption{(Color online)Plots of the entanglement entropy for various values of the number $Q$ of nodes on the closed dimension. From left to right and top to bottom: $Q=1,5,6,7$, respectively.}
\label{FigSNsmall} 
\end{figure}

In Fig. (\ref{FigSNsmall}) we have represented the asymptotic entropy of entanglement, obtained
from Eq. (\ref{eq:sumqi}), for some representative cases, as a function
of $\theta$ and $\phi$. In the asymptotic time behaviour, the entropy derived from (\ref{eq:limEnt}) is maximal for $\theta=\pi/2$ and $\phi=\frac{\pi}{2}$ or $\frac{3\pi}{2}$, with a corresponding value: 
\begin{equation}
S_{max}=\frac{2}{\pi}\log_{2}\frac{2}{\pi}+(1-\frac{2}{\pi})\log_{2}(1-\frac{2}{\pi})\simeq0.945.
\end{equation}
Such value clearly overcomes the corresponding limit with only one
dimension. The differences observed in the amount of entanglement
generated within the QW on a cylinder, as compared to the ordinary
QW, may have important consequences. The QW has been suggested as
a possible device to generate entanglement in quantum information
processes \cite{goyal2010spatial}. On the other hand, the coin can
be regarded as a thermodynamic subsystem interacting with the lattice.
As such, it becomes an interesting scenario to investigate the approach
to thermodynamical equilibrium in quantum systems \cite{romanelli2012thermodynamic}.
We have shown that the QW on a cylinder behaves differently to the
QW, with a dynamics that allows to reach larger values of entanglement.
Therefore, it is possible that the transition towards equilibrium
will show new features. Among these features is the investigation
of a non-Markovian behavior previous to the asymptotic regime, as
already observed for the QW \cite{hinarejos2014chirality}.

\section{The continuous limit\label{sec:IV}}

A practical tool to study the analytical properties of QWs on the
discrete circle is looking at the continuous limit of the DFT of the
walk \cite{di2012discrete}. In order to take the continuous limit
we introduce, in the unitary operator (\ref{eq:cycleU}), $k=\tilde{k}\Delta x$,
$q_{i}=\tilde{q}_{i}\Delta y$ and $t=j\Delta t$, where ($\Delta x$,
$\Delta y$, $\Delta t$) are the step sizes of the space-time lattice.
Then we introduce an infinitesimal $\epsilon$ and write $\Delta t=\Delta x=\Delta y=\epsilon$,
and assume that all functions are at least $C^{2}$-differentiable
in all their arguments. We now expand the original discrete equations,
defined by the unitary operator (\ref{eq:cycleU}) around $\epsilon=0$.
A necessary and sufficient condition for the expression to be self-consistent
at order $0$ in $\epsilon$ is that $\hat{U}_{q_{i}}(k)^{(0)}=\hat{C}_{y}\hat{C}_{x}=\mathbb{I}$.
This is satisfied in our AQW because we chose that angles $\alpha_i$ and $\beta_i$ are zero. If we Taylor expand each term around $\epsilon = 0$, the zeroth-order terms identically vanish and the next lowest order contribution in O($\epsilon$) recovers a couple of partial differential equations for the two-component wave function $\psi$. Notice
that the limit is taken on both dimensions, keeping $q_{i}$ constant. A tedious but straightforward computation gives the following equation in physical space, obeyed by the wave function $\psi$: 
\begin{equation}
\psi_{t}-\sigma_{z}\psi_{x}=iq_{i}\psi.\label{eq:dirac2}
\end{equation}
We observe that this couple of equations coincide with the massive (1+1)-Dirac equation, where the pseudo-momentum $q_{i}$ plays the role of the fermion mass term. This is consistent with the analysis of section \ref{sub:II_B}. Indeed, the dispersion relation in Eq. (\ref{eq:DR}) recovers the usual Dirac cone in continuous limit. More in particular this result shows that QWs on a cylinder can be used to model quantum transport of a fermion with a mass $m \in [0, q_{max}]$, where $q_{max}$ is the UV cut-off on the closed dimension and simulate fermions with different masses easily by a suitable choice of the initial condition. 

\section{Conclusions and outlook}

In this work we have analyzed a quantum walk defined on a cylinder.
A simple approach to this problem is given by the alternated use of
a single qubit on the two dimensions \cite{CdiF_PRL2011}. Although
the size of the closed $y$ dimension is in principle arbitrary, it
is in the spirit of this paper to regard this dimension as a sort
of ``hidden'' or ``extra'' dimension, i.e. it is defined by some
length scale which is much smaller that the observed spreading along
the open $x$ coordinate. The reason for this approach is twofold.
First, some physical devices, such as nanotubes, can be effectively
described in this way. Secondly, one can establish a connection with
theories in high energy physics that assume the existence of compactified
extra dimensions. In fact, the QW has been shown as a candidate to
simulate many physical phenomena, ranging from the motion of a particle
on a curved space-time \cite{di2013quantum} to Yang-Mills gauge theories
\cite{Arnault2016} or neutrino oscillations \cite{Mallick2016,Molfetta2016}.
In this
spirit, we analyzed the motion along the open coordinate after tracing
out the closed dimension. After this, the ``hidden dimension'' manifests
in providing several components to the observed QW, which arise from
different values $q_{i}$ of the quasi-momentum in the closed direction.
These components move each one with a different velocity, which originate
from the corresponding group velocity at a given $q_{i}$.

One can get more insight about the role played by the closed dimension
on the infinite one by examining the continuous limit of the QW, where
a particular value $q_{i}$ is selected. One then arrives to a Dirac equation
describing the motion along the open coordinate, where the pseudo-momentum
$q_{i}$ plays the role of the mass term. Therefore, by selecting
the appropriate quasi-momentum, one can easily simulate a family of
Dirac equations with different masses, a situation that reminds the
tower spectrum in Kaluza-Klein theories \cite{Witten1981,RubakovPhys.Usp.44:871-8932001;Usp.Fiz.Nauk171:913-9382001}.
In our opinion, the simulation of the QW on a cylinder opens the possibility
to investigate many aspects that appear in many high energy theories
and, thus, deserves further attention.

\section{Acknowledgements}
This work has been supported by the Spanish Ministerio de Educaci{\'o}n
e Innovaci{\'o}n, MICIN-FEDER projects FPA2014-54459-P, FIS2014-60715-P, SEV-2014-0398 and
Generalitat Valenciana grant GVPROMETEOII2014-087.

\appendix

\section{Entanglement entropy\label{sec:appendixA}}

We use the notation $\ket{\psi_{k,q_{i}}(j)}=\braket{k,q_{i}\mid\psi(j)}$,
which represents two-component spinor in the quasi-momentum basis.
With this notation, we can write 
\begin{equation}
\ket{\psi_{k,q_{i}}(0)}=\left(\begin{array}{c}
\cos\frac{\theta}{2}\\
e^{i\phi}\sin\frac{\theta}{2}
\end{array}\right)
\end{equation}

Using the unitary operator (\ref{eq:AEQWop}), for $\theta_{x}=\theta_{y}=\frac{\pi}{4}$,
in this basis, it follows that 
\begin{equation}
\ket{\psi_{k,q_{i}}(j)}=\hat{U}_{q_{i}}(k)\ket{\psi_{k,q_{i}}(0)},\label{eq:iterpsip}
\end{equation}
with 
\begin{equation}
\hat{U}_{q_{i}}(k)=\left(\begin{array}{cc}
e^{iq_{i}}\cos k & -ie^{iq_{i}}\sin k\\
-ie^{-iq_{i}}\sin k & e^{-iq_{i}}\cos k
\end{array}\right).\label{eq:cycleU}
\end{equation}

The $t$-th power of $\hat{U}_{q_{i}}(k)$ is obtained from the spectral
theorem: 
\begin{equation}
\hat{U}_{q_{i}}^{j}(k)=\sum_{h=\pm 1}e^{-i\omega_{h}(k,q_{i})j}\ket{\phi_{h}(k,q_{i})}\bra{\phi_{h}(k,q_{i})}.\label{eq:UDApt}
\end{equation}
In the latter equation, $\omega_{h}(k,q_{i})$ is obtained from the
dispersion equation, and $\ket{\phi_{h}(k,q_{i})},h=\pm 1$ are the
two normalized eigenvectors of $\hat{U}_{q_{i}}(k)$, given by: 
\begin{equation}
\ket{\phi_{h}(k,q_{i})}=\frac{1}{N_{h}}\left(\begin{array}{c}
e^{iq_{i}}\sin k\\
-\sin q_{i}\cos k\pm\sin\omega
\end{array}\right),\label{eq:eigenp}
\end{equation}
respectively for $h=\pm 1$, and $N_{h}$ is an appropriate normalization
constant. From the above expressions one can obtain $\rho_{c}(j)$
as 
\begin{equation}
\rho_{c}(j)=\frac{1}{Q}\sum_{i=0}^{Q-1}\int_{-\pi}^{\pi}\frac{dk}{2\pi}\ket{\psi_{k,q_{i}}(j)}\bra{\psi_{k,q_{i}}(j)}.
\end{equation}

Eq. (\ref{eq:UDApt}) contains terms of the form $e^{\pm2i\omega_{h}j}$.
For large values of $j$, such terms become highly oscillatory, while
the rest of terms that depend on the variables $k$ and $q_{i}$ are
smooth functions. We can therefore neglect the integral over $k$
of such strongly oscillatory terms. By doing so, we arrive to the
expression 
\begin{equation}
\overset{\sim}{\rho_{c}}\equiv\lim_{j\rightarrow\infty}\rho_{c}(j)=\frac{1}{Q}\sum_{i=0}^{Q-1}\left(\begin{array}{cc}
1-r_{22}(q_{i}) & r_{12}(q_{i})\\
r_{12}^{*}(q_{i}) & r_{22}(q_{i})
\end{array}\right),\label{eq:sumqi}
\end{equation}
where 
\begin{align}
r_{22}(q) & =\frac{\cos^{2}\frac{\theta}{2}\left[1-\nu(q)\right]+\sin^{2}\frac{\theta}{2}\left[\cos2q+\nu(q)\right]}{2\cos^{2}q}
\end{align}
\begin{align}
r_{12}(q) & =\frac{e^{-2iq}e^{i\phi}+e^{-i\phi}}{4\cos^{2}q}\left[1-\nu(q)\right]\sin\theta
\end{align}
with $\nu(q)\equiv\frac{\sqrt{1-\cos(2q)}}{\sqrt{2}}.$

As shown in Fig. (\ref{FigSNsmall}), this magnitude can present different shapes as the
value of $Q$ is changed. More importantly, we observe that one reaches
values close to unity for some angles. For larger values the shape
stabilizes and looks similar to the case with $Q=7$. In fact, one
can derive a closed expression for $\overset{\sim}{\rho_{c}}$ in
the limit $Q\rightarrow\infty$, by replacing the sum in Eq. (\ref{eq:sumqi})
by an integral over the continuous variable $q$, giving the final
expression 
\begin{widetext}
\begin{equation}
\lim_{Q\rightarrow\infty}\overset{\sim}{\rho_{c}}=\frac{1}{2\pi}\left(\begin{array}{cc}
\pi+(\pi-2)\cos\theta & \left[e^{-i\phi}+(\pi-3)e^{i\phi}\right]\sin\theta\\
\left[e^{i\phi}+(\pi-3)e^{-i\phi}\right]\sin\theta & \pi-(\pi-2)\cos\theta
\end{array}\right).
\label{eq:limEnt}
\end{equation}

\end{widetext}

\bibliographystyle{apsrev4-1}
\bibliography{bib}

%merlin.mbs apsrev4-1.bst 2010-07-25 4.21a (PWD, AO, DPC) hacked
%Control: key (0)
%Control: author (72) initials jnrlst
%Control: editor formatted (1) identically to author
%Control: production of article title (-1) disabled
%Control: page (0) single
%Control: year (1) truncated
%Control: production of eprint (0) enabled
\begin{thebibliography}{42}%
\makeatletter
\providecommand \@ifxundefined [1]{%
 \@ifx{#1\undefined}
}%
\providecommand \@ifnum [1]{%
 \ifnum #1\expandafter \@firstoftwo
 \else \expandafter \@secondoftwo
 \fi
}%
\providecommand \@ifx [1]{%
 \ifx #1\expandafter \@firstoftwo
 \else \expandafter \@secondoftwo
 \fi
}%
\providecommand \natexlab [1]{#1}%
\providecommand \enquote  [1]{``#1''}%
\providecommand \bibnamefont  [1]{#1}%
\providecommand \bibfnamefont [1]{#1}%
\providecommand \citenamefont [1]{#1}%
\providecommand \href@noop [0]{\@secondoftwo}%
\providecommand \href [0]{\begingroup \@sanitize@url \@href}%
\providecommand \@href[1]{\@@startlink{#1}\@@href}%
\providecommand \@@href[1]{\endgroup#1\@@endlink}%
\providecommand \@sanitize@url [0]{\catcode `\\12\catcode `\$12\catcode
  `\&12\catcode `\#12\catcode `\^12\catcode `\_12\catcode `\%12\relax}%
\providecommand \@@startlink[1]{}%
\providecommand \@@endlink[0]{}%
\providecommand \url  [0]{\begingroup\@sanitize@url \@url }%
\providecommand \@url [1]{\endgroup\@href {#1}{\urlprefix }}%
\providecommand \urlprefix  [0]{URL }%
\providecommand \Eprint [0]{\href }%
\providecommand \doibase [0]{http://dx.doi.org/}%
\providecommand \selectlanguage [0]{\@gobble}%
\providecommand \bibinfo  [0]{\@secondoftwo}%
\providecommand \bibfield  [0]{\@secondoftwo}%
\providecommand \translation [1]{[#1]}%
\providecommand \BibitemOpen [0]{}%
\providecommand \bibitemStop [0]{}%
\providecommand \bibitemNoStop [0]{.\EOS\space}%
\providecommand \EOS [0]{\spacefactor3000\relax}%
\providecommand \BibitemShut  [1]{\csname bibitem#1\endcsname}%
\let\auto@bib@innerbib\@empty
%</preamble>
\bibitem [{\citenamefont {Kempe}(2003)}]{kempe2003quantum}%
  \BibitemOpen
  \bibfield  {author} {\bibinfo {author} {\bibfnamefont {J.}~\bibnamefont
  {Kempe}},\ }\href@noop {} {\bibfield  {journal} {\bibinfo  {journal}
  {Contemporary Physics}\ }\textbf {\bibinfo {volume} {44}},\ \bibinfo {pages}
  {307} (\bibinfo {year} {2003})}\BibitemShut {NoStop}%
\bibitem [{\citenamefont {Kendon}(2006)}]{kendon2006quantum}%
  \BibitemOpen
  \bibfield  {author} {\bibinfo {author} {\bibfnamefont {V.}~\bibnamefont
  {Kendon}},\ }\href@noop {} {\bibfield  {journal} {\bibinfo  {journal}
  {International Journal of Quantum Information}\ }\textbf {\bibinfo {volume}
  {4}},\ \bibinfo {pages} {791} (\bibinfo {year} {2006})}\BibitemShut {NoStop}%
\bibitem [{\citenamefont {Konno}(2008)}]{konno2008quantum}%
  \BibitemOpen
  \bibfield  {author} {\bibinfo {author} {\bibfnamefont {N.}~\bibnamefont
  {Konno}},\ }in\ \href@noop {} {\emph {\bibinfo {booktitle} {Quantum Potential
  Theory}}}\ (\bibinfo  {publisher} {Springer},\ \bibinfo {year} {2008})\ pp.\
  \bibinfo {pages} {309--452}\BibitemShut {NoStop}%
\bibitem [{\citenamefont {Venegas-Andraca}(2012)}]{venegas2012quantum}%
  \BibitemOpen
  \bibfield  {author} {\bibinfo {author} {\bibfnamefont {S.~E.}\ \bibnamefont
  {Venegas-Andraca}},\ }\href@noop {} {\bibfield  {journal} {\bibinfo
  {journal} {Quantum Information Processing}\ }\textbf {\bibinfo {volume}
  {11}},\ \bibinfo {pages} {1015} (\bibinfo {year} {2012})}\BibitemShut
  {NoStop}%
\bibitem [{\citenamefont {M{\"u}lken}\ and\ \citenamefont
  {Blumen}(2011)}]{mulken2011continuous}%
  \BibitemOpen
  \bibfield  {author} {\bibinfo {author} {\bibfnamefont {O.}~\bibnamefont
  {M{\"u}lken}}\ and\ \bibinfo {author} {\bibfnamefont {A.}~\bibnamefont
  {Blumen}},\ }\href@noop {} {\bibfield  {journal} {\bibinfo  {journal}
  {Physics Reports}\ }\textbf {\bibinfo {volume} {502}},\ \bibinfo {pages} {37}
  (\bibinfo {year} {2011})}\BibitemShut {NoStop}%
\bibitem [{\citenamefont {Aharonov}\ \emph {et~al.}(1993)\citenamefont
  {Aharonov}, \citenamefont {Davidovich},\ and\ \citenamefont
  {Zagury}}]{aharonov1993quantum}%
  \BibitemOpen
  \bibfield  {author} {\bibinfo {author} {\bibfnamefont {Y.}~\bibnamefont
  {Aharonov}}, \bibinfo {author} {\bibfnamefont {L.}~\bibnamefont
  {Davidovich}}, \ and\ \bibinfo {author} {\bibfnamefont {N.}~\bibnamefont
  {Zagury}},\ }\href@noop {} {\bibfield  {journal} {\bibinfo  {journal}
  {Physical Review A}\ }\textbf {\bibinfo {volume} {48}},\ \bibinfo {pages}
  {1687} (\bibinfo {year} {1993})}\BibitemShut {NoStop}%
\bibitem [{\citenamefont {Ambainis}\ \emph {et~al.}(2001)\citenamefont
  {Ambainis}, \citenamefont {Bach}, \citenamefont {Nayak}, \citenamefont
  {Vishwanath},\ and\ \citenamefont {Watrous}}]{ambainis2001one}%
  \BibitemOpen
  \bibfield  {author} {\bibinfo {author} {\bibfnamefont {A.}~\bibnamefont
  {Ambainis}}, \bibinfo {author} {\bibfnamefont {E.}~\bibnamefont {Bach}},
  \bibinfo {author} {\bibfnamefont {A.}~\bibnamefont {Nayak}}, \bibinfo
  {author} {\bibfnamefont {A.}~\bibnamefont {Vishwanath}}, \ and\ \bibinfo
  {author} {\bibfnamefont {J.}~\bibnamefont {Watrous}},\ }in\ \href@noop {}
  {\emph {\bibinfo {booktitle} {Proceedings of the thirty-third annual ACM
  symposium on Theory of computing}}}\ (\bibinfo {organization} {ACM},\
  \bibinfo {year} {2001})\ pp.\ \bibinfo {pages} {37--49}\BibitemShut {NoStop}%
\bibitem [{\citenamefont {Childs}(2009)}]{childs2009universal}%
  \BibitemOpen
  \bibfield  {author} {\bibinfo {author} {\bibfnamefont {A.~M.}\ \bibnamefont
  {Childs}},\ }\href@noop {} {\bibfield  {journal} {\bibinfo  {journal}
  {Physical Review Letters}\ }\textbf {\bibinfo {volume} {102}},\ \bibinfo
  {pages} {180501} (\bibinfo {year} {2009})}\BibitemShut {NoStop}%
\bibitem [{\citenamefont {Lovett}\ \emph {et~al.}(2010)\citenamefont {Lovett},
  \citenamefont {Cooper}, \citenamefont {Everitt}, \citenamefont {Trevers},\
  and\ \citenamefont {Kendon}}]{lovett2010universal}%
  \BibitemOpen
  \bibfield  {author} {\bibinfo {author} {\bibfnamefont {N.~B.}\ \bibnamefont
  {Lovett}}, \bibinfo {author} {\bibfnamefont {S.}~\bibnamefont {Cooper}},
  \bibinfo {author} {\bibfnamefont {M.}~\bibnamefont {Everitt}}, \bibinfo
  {author} {\bibfnamefont {M.}~\bibnamefont {Trevers}}, \ and\ \bibinfo
  {author} {\bibfnamefont {V.}~\bibnamefont {Kendon}},\ }\href@noop {}
  {\bibfield  {journal} {\bibinfo  {journal} {Physical Review A}\ }\textbf
  {\bibinfo {volume} {81}},\ \bibinfo {pages} {042330} (\bibinfo {year}
  {2010})}\BibitemShut {NoStop}%
\bibitem [{\citenamefont {de~Valc{\'a}rcel}\ \emph {et~al.}(2010)\citenamefont
  {de~Valc{\'a}rcel}, \citenamefont {Rold{\'a}n},\ and\ \citenamefont
  {Romanelli}}]{GdeValc_NJP2010}%
  \BibitemOpen
  \bibfield  {author} {\bibinfo {author} {\bibfnamefont {G.~J.}\ \bibnamefont
  {de~Valc{\'a}rcel}}, \bibinfo {author} {\bibfnamefont {E.}~\bibnamefont
  {Rold{\'a}n}}, \ and\ \bibinfo {author} {\bibfnamefont {A.}~\bibnamefont
  {Romanelli}},\ }\href {http://stacks.iop.org/1367-2630/12/i=12/a=123022}
  {\bibfield  {journal} {\bibinfo  {journal} {New Journal of Physics}\ }\textbf
  {\bibinfo {volume} {12}},\ \bibinfo {pages} {123022} (\bibinfo {year}
  {2010})}\BibitemShut {NoStop}%
\bibitem [{\citenamefont {Hinarejos}\ \emph {et~al.}(2013)\citenamefont
  {Hinarejos}, \citenamefont {P{\'e}rez}, \citenamefont {Rold{\`a}n},
  \citenamefont {Romanelli},\ and\ \citenamefont
  {de~Valcarcel}}]{hinarejos2013understanding}%
  \BibitemOpen
  \bibfield  {author} {\bibinfo {author} {\bibfnamefont {M.}~\bibnamefont
  {Hinarejos}}, \bibinfo {author} {\bibfnamefont {A.}~\bibnamefont
  {P{\'e}rez}}, \bibinfo {author} {\bibfnamefont {E.}~\bibnamefont
  {Rold{\`a}n}}, \bibinfo {author} {\bibfnamefont {A.}~\bibnamefont
  {Romanelli}}, \ and\ \bibinfo {author} {\bibfnamefont {G.}~\bibnamefont
  {de~Valcarcel}},\ }\href@noop {} {\bibfield  {journal} {\bibinfo  {journal}
  {New Journal of Physics}\ }\textbf {\bibinfo {volume} {15}},\ \bibinfo
  {pages} {073041} (\bibinfo {year} {2013})}\BibitemShut {NoStop}%
\bibitem [{\citenamefont {Strauch}(2006)}]{strauch2006relativistic}%
  \BibitemOpen
  \bibfield  {author} {\bibinfo {author} {\bibfnamefont {F.~W.}\ \bibnamefont
  {Strauch}},\ }\href@noop {} {\bibfield  {journal} {\bibinfo  {journal}
  {Physical Review A}\ }\textbf {\bibinfo {volume} {73}},\ \bibinfo {pages}
  {054302} (\bibinfo {year} {2006})}\BibitemShut {NoStop}%
\bibitem [{\citenamefont {Di~Molfetta}\ and\ \citenamefont
  {Debbasch}(2012)}]{di2012discrete}%
  \BibitemOpen
  \bibfield  {author} {\bibinfo {author} {\bibfnamefont {G.}~\bibnamefont
  {Di~Molfetta}}\ and\ \bibinfo {author} {\bibfnamefont {F.}~\bibnamefont
  {Debbasch}},\ }\href@noop {} {\bibfield  {journal} {\bibinfo  {journal}
  {Journal of Mathematical Physics}\ }\textbf {\bibinfo {volume} {53}},\
  \bibinfo {pages} {123302} (\bibinfo {year} {2012})}\BibitemShut {NoStop}%
\bibitem [{\citenamefont {Di~Molfetta}\ \emph {et~al.}(2013)\citenamefont
  {Di~Molfetta}, \citenamefont {Brachet},\ and\ \citenamefont
  {Debbasch}}]{di2013quantum}%
  \BibitemOpen
  \bibfield  {author} {\bibinfo {author} {\bibfnamefont {G.}~\bibnamefont
  {Di~Molfetta}}, \bibinfo {author} {\bibfnamefont {M.}~\bibnamefont
  {Brachet}}, \ and\ \bibinfo {author} {\bibfnamefont {F.}~\bibnamefont
  {Debbasch}},\ }\href@noop {} {\bibfield  {journal} {\bibinfo  {journal}
  {Physical Review A}\ }\textbf {\bibinfo {volume} {88}},\ \bibinfo {pages}
  {042301} (\bibinfo {year} {2013})}\BibitemShut {NoStop}%
\bibitem [{\citenamefont {Manouchehri}\ and\ \citenamefont
  {Wang}(2013)}]{Manouchehri:2013:PIQ:2566741}%
  \BibitemOpen
  \bibfield  {author} {\bibinfo {author} {\bibfnamefont {K.}~\bibnamefont
  {Manouchehri}}\ and\ \bibinfo {author} {\bibfnamefont {J.}~\bibnamefont
  {Wang}},\ }\href@noop {} {\emph {\bibinfo {title} {Physical Implementation of
  Quantum Walks}}}\ (\bibinfo  {publisher} {Springer Publishing Company,
  Incorporated},\ \bibinfo {year} {2013})\BibitemShut {NoStop}%
\bibitem [{\citenamefont {Preiss}\ \emph {et~al.}(2015)\citenamefont {Preiss},
  \citenamefont {Ma}, \citenamefont {Tai}, \citenamefont {Lukin}, \citenamefont
  {Rispoli}, \citenamefont {Zupancic}, \citenamefont {Lahini}, \citenamefont
  {Islam},\ and\ \citenamefont {Greiner}}]{preiss2015strongly}%
  \BibitemOpen
  \bibfield  {author} {\bibinfo {author} {\bibfnamefont {P.~M.}\ \bibnamefont
  {Preiss}}, \bibinfo {author} {\bibfnamefont {R.}~\bibnamefont {Ma}}, \bibinfo
  {author} {\bibfnamefont {M.~E.}\ \bibnamefont {Tai}}, \bibinfo {author}
  {\bibfnamefont {A.}~\bibnamefont {Lukin}}, \bibinfo {author} {\bibfnamefont
  {M.}~\bibnamefont {Rispoli}}, \bibinfo {author} {\bibfnamefont
  {P.}~\bibnamefont {Zupancic}}, \bibinfo {author} {\bibfnamefont
  {Y.}~\bibnamefont {Lahini}}, \bibinfo {author} {\bibfnamefont
  {R.}~\bibnamefont {Islam}}, \ and\ \bibinfo {author} {\bibfnamefont
  {M.}~\bibnamefont {Greiner}},\ }\href@noop {} {\bibfield  {journal} {\bibinfo
   {journal} {Science}\ }\textbf {\bibinfo {volume} {347}},\ \bibinfo {pages}
  {1229} (\bibinfo {year} {2015})}\BibitemShut {NoStop}%
\bibitem [{\citenamefont {Matsue}\ \emph {et~al.}(2016)\citenamefont {Matsue},
  \citenamefont {Ogurisu},\ and\ \citenamefont {Segawa}}]{matsue2016quantum}%
  \BibitemOpen
  \bibfield  {author} {\bibinfo {author} {\bibfnamefont {K.}~\bibnamefont
  {Matsue}}, \bibinfo {author} {\bibfnamefont {O.}~\bibnamefont {Ogurisu}}, \
  and\ \bibinfo {author} {\bibfnamefont {E.}~\bibnamefont {Segawa}},\
  }\href@noop {} {\bibfield  {journal} {\bibinfo  {journal} {Quantum
  Information Processing}\ }\textbf {\bibinfo {volume} {15}},\ \bibinfo {pages}
  {1865} (\bibinfo {year} {2016})}\BibitemShut {NoStop}%
\bibitem [{\citenamefont {Rold\'an}\ \emph {et~al.}(2013)\citenamefont
  {Rold\'an}, \citenamefont {Di~Franco}, \citenamefont {Silva},\ and\
  \citenamefont {de~Valc\'arcel}}]{roldan2013n}%
  \BibitemOpen
  \bibfield  {author} {\bibinfo {author} {\bibfnamefont {E.}~\bibnamefont
  {Rold\'an}}, \bibinfo {author} {\bibfnamefont {C.}~\bibnamefont {Di~Franco}},
  \bibinfo {author} {\bibfnamefont {F.}~\bibnamefont {Silva}}, \ and\ \bibinfo
  {author} {\bibfnamefont {G.~J.}\ \bibnamefont {de~Valc\'arcel}},\ }\href
  {\doibase 10.1103/PhysRevA.87.022336} {\bibfield  {journal} {\bibinfo
  {journal} {Phys. Rev. A}\ }\textbf {\bibinfo {volume} {87}},\ \bibinfo
  {pages} {022336} (\bibinfo {year} {2013})}\BibitemShut {NoStop}%
\bibitem [{\citenamefont {Knight}\ \emph {et~al.}(2004)\citenamefont {Knight},
  \citenamefont {Rold{\'a}n},\ and\ \citenamefont
  {Sipe}}]{knight2004propagating}%
  \BibitemOpen
  \bibfield  {author} {\bibinfo {author} {\bibfnamefont {P.~L.}\ \bibnamefont
  {Knight}}, \bibinfo {author} {\bibfnamefont {E.}~\bibnamefont {Rold{\'a}n}},
  \ and\ \bibinfo {author} {\bibfnamefont {J.}~\bibnamefont {Sipe}},\
  }\href@noop {} {\bibfield  {journal} {\bibinfo  {journal} {Journal of Modern
  Optics}\ }\textbf {\bibinfo {volume} {51}},\ \bibinfo {pages} {1761}
  (\bibinfo {year} {2004})}\BibitemShut {NoStop}%
\bibitem [{\citenamefont {Witten}(1981)}]{Witten1981}%
  \BibitemOpen
  \bibfield  {author} {\bibinfo {author} {\bibfnamefont {E.}~\bibnamefont
  {Witten}},\ }\href {\doibase http://dx.doi.org/10.1016/0550-3213(81)90021-3}
  {\bibfield  {journal} {\bibinfo  {journal} {Nuclear Physics B}\ }\textbf
  {\bibinfo {volume} {186}},\ \bibinfo {pages} {412 } (\bibinfo {year}
  {1981})}\BibitemShut {NoStop}%
\bibitem [{\citenamefont
  {Rubakov}(2001)}]{RubakovPhys.Usp.44:871-8932001;Usp.Fiz.Nauk171:913-9382001}%
  \BibitemOpen
  \bibfield  {author} {\bibinfo {author} {\bibfnamefont {V.~A.}\ \bibnamefont
  {Rubakov}},\ }\href@noop {} {\bibfield  {journal} {\bibinfo  {journal}
  {Physics-Uspekhi}\ }\textbf {\bibinfo {volume} {44}},\ \bibinfo {pages} {871}
  (\bibinfo {year} {2001})}\BibitemShut {NoStop}%
\bibitem [{\citenamefont {Ambainis}\ \emph {et~al.}(2005)\citenamefont
  {Ambainis}, \citenamefont {Kempe},\ and\ \citenamefont
  {Rivosh}}]{Ambainis:2005}%
  \BibitemOpen
  \bibfield  {author} {\bibinfo {author} {\bibfnamefont {A.}~\bibnamefont
  {Ambainis}}, \bibinfo {author} {\bibfnamefont {J.}~\bibnamefont {Kempe}}, \
  and\ \bibinfo {author} {\bibfnamefont {A.}~\bibnamefont {Rivosh}},\ }in\
  \href {http://dl.acm.org/citation.cfm?id=1070432.1070590} {\emph {\bibinfo
  {booktitle} {Proceedings of the Sixteenth Annual ACM-SIAM Symposium on
  Discrete Algorithms}}},\ \bibinfo {series and number} {SODA '05}\ (\bibinfo
  {publisher} {Society for Industrial and Applied Mathematics},\ \bibinfo
  {address} {Philadelphia, PA, USA},\ \bibinfo {year} {2005})\ pp.\ \bibinfo
  {pages} {1099--1108}\BibitemShut {NoStop}%
\bibitem [{\citenamefont {Di~Franco}\ \emph
  {et~al.}(2011{\natexlab{a}})\citenamefont {Di~Franco}, \citenamefont
  {Mc~Gettrick},\ and\ \citenamefont {Busch}}]{CdiF_PRL2011}%
  \BibitemOpen
  \bibfield  {author} {\bibinfo {author} {\bibfnamefont {C.}~\bibnamefont
  {Di~Franco}}, \bibinfo {author} {\bibfnamefont {M.}~\bibnamefont
  {Mc~Gettrick}}, \ and\ \bibinfo {author} {\bibfnamefont {T.}~\bibnamefont
  {Busch}},\ }\href {\doibase 10.1103/PhysRevLett.106.080502} {\bibfield
  {journal} {\bibinfo  {journal} {Phys. Rev. Lett.}\ }\textbf {\bibinfo
  {volume} {106}},\ \bibinfo {pages} {080502} (\bibinfo {year}
  {2011}{\natexlab{a}})}\BibitemShut {NoStop}%
\bibitem [{\citenamefont {Di~Franco}\ \emph
  {et~al.}(2011{\natexlab{b}})\citenamefont {Di~Franco}, \citenamefont
  {Mc~Gettrick}, \citenamefont {Machida},\ and\ \citenamefont
  {Busch}}]{CdiF_PRA2011}%
  \BibitemOpen
  \bibfield  {author} {\bibinfo {author} {\bibfnamefont {C.}~\bibnamefont
  {Di~Franco}}, \bibinfo {author} {\bibfnamefont {M.}~\bibnamefont
  {Mc~Gettrick}}, \bibinfo {author} {\bibfnamefont {T.}~\bibnamefont
  {Machida}}, \ and\ \bibinfo {author} {\bibfnamefont {T.}~\bibnamefont
  {Busch}},\ }\href {\doibase 10.1103/PhysRevA.84.042337} {\bibfield  {journal}
  {\bibinfo  {journal} {Phys. Rev. A}\ }\textbf {\bibinfo {volume} {84}},\
  \bibinfo {pages} {042337} (\bibinfo {year} {2011}{\natexlab{b}})}\BibitemShut
  {NoStop}%
\bibitem [{\citenamefont {Nayak}\ and\ \citenamefont
  {Vishwanath}(2007)}]{Nayak2007}%
  \BibitemOpen
  \bibfield  {author} {\bibinfo {author} {\bibfnamefont {A.}~\bibnamefont
  {Nayak}}\ and\ \bibinfo {author} {\bibfnamefont {A.}~\bibnamefont
  {Vishwanath}},\ }\href@noop {} {\  (\bibinfo {year} {2007})},\ \Eprint
  {http://arxiv.org/abs/quant-ph/0010117} {quant-ph/0010117} \BibitemShut
  {NoStop}%
\bibitem [{\citenamefont {Bru}\ \emph {et~al.}(2016)\citenamefont {Bru},
  \citenamefont {Hinarejos}, \citenamefont {Silva}, \citenamefont
  {de~Valc\'arcel},\ and\ \citenamefont {Rold\'an}}]{LBru_PRA2016}%
  \BibitemOpen
  \bibfield  {author} {\bibinfo {author} {\bibfnamefont {L.~A.}\ \bibnamefont
  {Bru}}, \bibinfo {author} {\bibfnamefont {M.}~\bibnamefont {Hinarejos}},
  \bibinfo {author} {\bibfnamefont {F.}~\bibnamefont {Silva}}, \bibinfo
  {author} {\bibfnamefont {G.~J.}\ \bibnamefont {de~Valc\'arcel}}, \ and\
  \bibinfo {author} {\bibfnamefont {E.}~\bibnamefont {Rold\'an}},\ }\href
  {\doibase 10.1103/PhysRevA.93.032333} {\bibfield  {journal} {\bibinfo
  {journal} {Phys. Rev. A}\ }\textbf {\bibinfo {volume} {93}},\ \bibinfo
  {pages} {032333} (\bibinfo {year} {2016})}\BibitemShut {NoStop}%
\bibitem [{\citenamefont {Carneiro}\ \emph {et~al.}(2005)\citenamefont
  {Carneiro}, \citenamefont {Loo}, \citenamefont {Xu}, \citenamefont {Girerd},
  \citenamefont {Kendon},\ and\ \citenamefont
  {Knight}}]{carneiro2005entanglement}%
  \BibitemOpen
  \bibfield  {author} {\bibinfo {author} {\bibfnamefont {I.}~\bibnamefont
  {Carneiro}}, \bibinfo {author} {\bibfnamefont {M.}~\bibnamefont {Loo}},
  \bibinfo {author} {\bibfnamefont {X.}~\bibnamefont {Xu}}, \bibinfo {author}
  {\bibfnamefont {M.}~\bibnamefont {Girerd}}, \bibinfo {author} {\bibfnamefont
  {V.}~\bibnamefont {Kendon}}, \ and\ \bibinfo {author} {\bibfnamefont {P.~L.}\
  \bibnamefont {Knight}},\ }\href@noop {} {\bibfield  {journal} {\bibinfo
  {journal} {New Journal of Physics}\ }\textbf {\bibinfo {volume} {7}},\
  \bibinfo {pages} {156} (\bibinfo {year} {2005})}\BibitemShut {NoStop}%
\bibitem [{\citenamefont {Venegas-Andraca}\ \emph {et~al.}(2005)\citenamefont
  {Venegas-Andraca}, \citenamefont {Ball}, \citenamefont {Burnett},\ and\
  \citenamefont {Bose}}]{venegas2005quantum}%
  \BibitemOpen
  \bibfield  {author} {\bibinfo {author} {\bibfnamefont {S.}~\bibnamefont
  {Venegas-Andraca}}, \bibinfo {author} {\bibfnamefont {J.}~\bibnamefont
  {Ball}}, \bibinfo {author} {\bibfnamefont {K.}~\bibnamefont {Burnett}}, \
  and\ \bibinfo {author} {\bibfnamefont {S.}~\bibnamefont {Bose}},\ }\href@noop
  {} {\bibfield  {journal} {\bibinfo  {journal} {New Journal of Physics}\
  }\textbf {\bibinfo {volume} {7}},\ \bibinfo {pages} {221} (\bibinfo {year}
  {2005})}\BibitemShut {NoStop}%
\bibitem [{\citenamefont {Endrejat}\ and\ \citenamefont
  {Buettner}(2005)}]{endrejat2005entanglement}%
  \BibitemOpen
  \bibfield  {author} {\bibinfo {author} {\bibfnamefont {J.}~\bibnamefont
  {Endrejat}}\ and\ \bibinfo {author} {\bibfnamefont {H.}~\bibnamefont
  {Buettner}},\ }\href@noop {} {\bibfield  {journal} {\bibinfo  {journal}
  {Journal of Physics A: Mathematical and General}\ }\textbf {\bibinfo {volume}
  {38}},\ \bibinfo {pages} {9289} (\bibinfo {year} {2005})}\BibitemShut
  {NoStop}%
\bibitem [{\citenamefont {Abal}\ \emph {et~al.}(2006)\citenamefont {Abal},
  \citenamefont {Siri}, \citenamefont {Romanelli},\ and\ \citenamefont
  {Donangelo}}]{abal2006quantum}%
  \BibitemOpen
  \bibfield  {author} {\bibinfo {author} {\bibfnamefont {G.}~\bibnamefont
  {Abal}}, \bibinfo {author} {\bibfnamefont {R.}~\bibnamefont {Siri}}, \bibinfo
  {author} {\bibfnamefont {A.}~\bibnamefont {Romanelli}}, \ and\ \bibinfo
  {author} {\bibfnamefont {R.}~\bibnamefont {Donangelo}},\ }\href@noop {}
  {\bibfield  {journal} {\bibinfo  {journal} {Physical Review A}\ }\textbf
  {\bibinfo {volume} {73}},\ \bibinfo {pages} {042302} (\bibinfo {year}
  {2006})}\BibitemShut {NoStop}%
\bibitem [{\citenamefont {Omar}\ \emph {et~al.}(2006)\citenamefont {Omar},
  \citenamefont {Paunkovi{\'c}}, \citenamefont {Sheridan},\ and\ \citenamefont
  {Bose}}]{omar2006quantum}%
  \BibitemOpen
  \bibfield  {author} {\bibinfo {author} {\bibfnamefont {Y.}~\bibnamefont
  {Omar}}, \bibinfo {author} {\bibfnamefont {N.}~\bibnamefont {Paunkovi{\'c}}},
  \bibinfo {author} {\bibfnamefont {L.}~\bibnamefont {Sheridan}}, \ and\
  \bibinfo {author} {\bibfnamefont {S.}~\bibnamefont {Bose}},\ }\href@noop {}
  {\bibfield  {journal} {\bibinfo  {journal} {Physical Review A}\ }\textbf
  {\bibinfo {volume} {74}},\ \bibinfo {pages} {042304} (\bibinfo {year}
  {2006})}\BibitemShut {NoStop}%
\bibitem [{\citenamefont {Maloyer}\ and\ \citenamefont
  {Kendon}(2007)}]{maloyer2007decoherence}%
  \BibitemOpen
  \bibfield  {author} {\bibinfo {author} {\bibfnamefont {O.}~\bibnamefont
  {Maloyer}}\ and\ \bibinfo {author} {\bibfnamefont {V.}~\bibnamefont
  {Kendon}},\ }\href@noop {} {\bibfield  {journal} {\bibinfo  {journal} {New
  Journal of Physics}\ }\textbf {\bibinfo {volume} {9}},\ \bibinfo {pages} {87}
  (\bibinfo {year} {2007})}\BibitemShut {NoStop}%
\bibitem [{\citenamefont {Pathak}\ and\ \citenamefont
  {Agarwal}(2007)}]{pathak2007quantum}%
  \BibitemOpen
  \bibfield  {author} {\bibinfo {author} {\bibfnamefont {P.}~\bibnamefont
  {Pathak}}\ and\ \bibinfo {author} {\bibfnamefont {G.}~\bibnamefont
  {Agarwal}},\ }\href@noop {} {\bibfield  {journal} {\bibinfo  {journal}
  {Physical Review A}\ }\textbf {\bibinfo {volume} {75}},\ \bibinfo {pages}
  {032351} (\bibinfo {year} {2007})}\BibitemShut {NoStop}%
\bibitem [{\citenamefont {Liu}\ and\ \citenamefont
  {Petulante}(2009)}]{liu2009one}%
  \BibitemOpen
  \bibfield  {author} {\bibinfo {author} {\bibfnamefont {C.}~\bibnamefont
  {Liu}}\ and\ \bibinfo {author} {\bibfnamefont {N.}~\bibnamefont
  {Petulante}},\ }\href@noop {} {\bibfield  {journal} {\bibinfo  {journal}
  {Physical Review A}\ }\textbf {\bibinfo {volume} {79}},\ \bibinfo {pages}
  {032312} (\bibinfo {year} {2009})}\BibitemShut {NoStop}%
\bibitem [{\citenamefont {Annabestani}\ \emph {et~al.}(2010)\citenamefont
  {Annabestani}, \citenamefont {Abolhasani},\ and\ \citenamefont
  {Abal}}]{annabestani2010asymptotic}%
  \BibitemOpen
  \bibfield  {author} {\bibinfo {author} {\bibfnamefont {M.}~\bibnamefont
  {Annabestani}}, \bibinfo {author} {\bibfnamefont {M.~R.}\ \bibnamefont
  {Abolhasani}}, \ and\ \bibinfo {author} {\bibfnamefont {G.}~\bibnamefont
  {Abal}},\ }\href@noop {} {\bibfield  {journal} {\bibinfo  {journal} {Journal
  of Physics A: Mathematical and Theoretical}\ }\textbf {\bibinfo {volume}
  {43}},\ \bibinfo {pages} {075301} (\bibinfo {year} {2010})}\BibitemShut
  {NoStop}%
\bibitem [{\citenamefont {de~Valc{\'a}rcel}\ \emph {et~al.}(2013)\citenamefont
  {de~Valc{\'a}rcel}, \citenamefont {Di~Franco}, \citenamefont {Hinarejos},
  \citenamefont {P{\'e}rez}, \citenamefont {Rold{\'a}n}, \citenamefont
  {Romanelli},\ and\ \citenamefont {Silva}}]{de2013multidimensional}%
  \BibitemOpen
  \bibfield  {author} {\bibinfo {author} {\bibfnamefont {G.}~\bibnamefont
  {de~Valc{\'a}rcel}}, \bibinfo {author} {\bibfnamefont {C.}~\bibnamefont
  {Di~Franco}}, \bibinfo {author} {\bibfnamefont {M.}~\bibnamefont
  {Hinarejos}}, \bibinfo {author} {\bibfnamefont {A.}~\bibnamefont
  {P{\'e}rez}}, \bibinfo {author} {\bibfnamefont {E.}~\bibnamefont
  {Rold{\'a}n}}, \bibinfo {author} {\bibfnamefont {A.}~\bibnamefont
  {Romanelli}}, \ and\ \bibinfo {author} {\bibfnamefont {F.}~\bibnamefont
  {Silva}},\ }in\ \href@noop {} {\emph {\bibinfo {booktitle} {2013 Conference
  on Lasers \& Electro-Optics Europe \& International Quantum Electronics
  Conference CLEO EUROPE/IQEC}}}\ (\bibinfo {year} {2013})\BibitemShut
  {NoStop}%
\bibitem [{\citenamefont {Goyal}\ and\ \citenamefont
  {Chandrashekar}(2010)}]{goyal2010spatial}%
  \BibitemOpen
  \bibfield  {author} {\bibinfo {author} {\bibfnamefont {S.~K.}\ \bibnamefont
  {Goyal}}\ and\ \bibinfo {author} {\bibfnamefont {C.}~\bibnamefont
  {Chandrashekar}},\ }\href@noop {} {\bibfield  {journal} {\bibinfo  {journal}
  {Journal of Physics A: Mathematical and Theoretical}\ }\textbf {\bibinfo
  {volume} {43}},\ \bibinfo {pages} {235303} (\bibinfo {year}
  {2010})}\BibitemShut {NoStop}%
\bibitem [{\citenamefont {Romanelli}(2012)}]{romanelli2012thermodynamic}%
  \BibitemOpen
  \bibfield  {author} {\bibinfo {author} {\bibfnamefont {A.}~\bibnamefont
  {Romanelli}},\ }\href@noop {} {\bibfield  {journal} {\bibinfo  {journal}
  {Physical Review A}\ }\textbf {\bibinfo {volume} {85}},\ \bibinfo {pages}
  {012319} (\bibinfo {year} {2012})}\BibitemShut {NoStop}%
\bibitem [{\citenamefont {Hinarejos}\ \emph {et~al.}(2014)\citenamefont
  {Hinarejos}, \citenamefont {Di~Franco}, \citenamefont {Romanelli},\ and\
  \citenamefont {P{\'e}rez}}]{hinarejos2014chirality}%
  \BibitemOpen
  \bibfield  {author} {\bibinfo {author} {\bibfnamefont {M.}~\bibnamefont
  {Hinarejos}}, \bibinfo {author} {\bibfnamefont {C.}~\bibnamefont
  {Di~Franco}}, \bibinfo {author} {\bibfnamefont {A.}~\bibnamefont
  {Romanelli}}, \ and\ \bibinfo {author} {\bibfnamefont {A.}~\bibnamefont
  {P{\'e}rez}},\ }\href@noop {} {\bibfield  {journal} {\bibinfo  {journal}
  {Physical Review A}\ }\textbf {\bibinfo {volume} {89}},\ \bibinfo {pages}
  {052330} (\bibinfo {year} {2014})}\BibitemShut {NoStop}%
\bibitem [{\citenamefont {Arnault}\ \emph {et~al.}(2016)\citenamefont
  {Arnault}, \citenamefont {Molfetta}, \citenamefont {Brachet},\ and\
  \citenamefont {Debbasch}}]{Arnault2016}%
  \BibitemOpen
  \bibfield  {author} {\bibinfo {author} {\bibfnamefont {P.}~\bibnamefont
  {Arnault}}, \bibinfo {author} {\bibfnamefont {G.~D.}\ \bibnamefont
  {Molfetta}}, \bibinfo {author} {\bibfnamefont {M.}~\bibnamefont {Brachet}}, \
  and\ \bibinfo {author} {\bibfnamefont {F.}~\bibnamefont {Debbasch}},\
  }\href@noop {} {\  (\bibinfo {year} {2016})},\ \Eprint
  {http://arxiv.org/abs/1605.01605} {1605.01605} \BibitemShut {NoStop}%
\bibitem [{\citenamefont {Mallick}\ \emph {et~al.}(2016)\citenamefont
  {Mallick}, \citenamefont {Mandal},\ and\ \citenamefont
  {Chandrashekar}}]{Mallick2016}%
  \BibitemOpen
  \bibfield  {author} {\bibinfo {author} {\bibfnamefont {A.}~\bibnamefont
  {Mallick}}, \bibinfo {author} {\bibfnamefont {S.}~\bibnamefont {Mandal}}, \
  and\ \bibinfo {author} {\bibfnamefont {C.~M.}\ \bibnamefont
  {Chandrashekar}},\ }\href@noop {} {\  (\bibinfo {year} {2016})},\ \Eprint
  {http://arxiv.org/abs/1604.04233} {1604.04233} \BibitemShut {NoStop}%
\bibitem [{\citenamefont {Di~Molfetta}\ and\ \citenamefont
  {Pérez}(2016)}]{Molfetta2016}%
  \BibitemOpen
  \bibfield  {author} {\bibinfo {author} {\bibfnamefont {G.}~\bibnamefont
  {Di~Molfetta}}\ and\ \bibinfo {author} {\bibfnamefont {A.}~\bibnamefont
  {Pérez}},\ }\href@noop {} {\  (\bibinfo {year} {2016})},\ \Eprint
  {http://arxiv.org/abs/ArXiv 1607.00529} {ArXiv 1607.00529} \BibitemShut
  {NoStop}%
\end{thebibliography}%

\end{document}